\documentclass[twocolumn,secnumarabic, amssymb, amsmath,tightenlines,showpacs,superscriptaddress,preprintnumbers, aps,prb]{revtex4-1}

\usepackage{times}
\usepackage{color}
\usepackage{graphicx}
\usepackage{dcolumn}
\usepackage{amssymb}
\usepackage{amsmath}
\usepackage{amsfonts}
\usepackage{docs}%
\usepackage{bm}%
\usepackage[colorlinks=true,linkcolor=blue,pagecolor=blue,filecolor=blue,menucolor=blue,urlcolor=blue,citecolor=blue,anchorcolor=blue]{hyperref}%

\newcommand{\etal}{\emph{et al.}}

\begin{document}

\title{Spin Controlled Coexistence of $0$ and $\pi$ States in $SFSFS$ Josephson Junctions }
\author{Mohammad Alidoust }
\email{phymalidoust@gmail.com}
\affiliation{Department of Physics,
Faculty of Sciences, University of Isfahan, Hezar Jerib Avenue,
Isfahan 81746-73441, Iran}
\author{Klaus Halterman}
\email{klaus.halterman@navy.mil} \affiliation{Michelson Lab, Physics
Division, Naval Air Warfare Center, China Lake, California 93555,
USA}

\date{\today}

\begin{abstract}
Using the Keldysh-Usadel formalism, we theoretically study the
$0$-$\pi$ transition profiles and current-phase relations of
magnetic $SFSFS$ and $SFSFFS$ Josephson nanojunctions in the
diffusive regime. By allowing the magnetizations of the
ferromagnetic layers to take arbitrary orientations, the strength
and direction of the charge supercurrent flowing through the
ferromagnetic regions can be controlled via the magnetization
rotation in one of the ferromagnetic layers. Depending on the
junction parameters, we find opposite current flow in the
ferromagnetic layers, revealing that remarkably such configurations
possess well-controlled $0$- and $\pi$-states simultaneously,
creating a three-terminal $0$-$\pi$ spin switch. We demonstrate that
the spin-controlled $0$-$\pi$ profiles trace back to the proximity
induced odd-frequency superconducting correlations generated by the
ferromagnetic layers. It is also shown that the spin-switching
effect can be more pronounced in $SFSFFS$ structures. The
current-phase relations reveal the important role of the middle $S$
electrode, where the spin controlled supercurrent depends crucially
on its thickness and phase differences with the outer $S$ terminals.
\end{abstract}

\pacs{74.50.+r, 74.45.+c, 74.25.Ha, 74.78.Na}

\maketitle

\section{Introduction}
It has been over a decade since hybrid structures of ferromagnets
and superconductors began to attract considerable interest from a
fundamental physics perspective as well as from the viewpoint of
practical devices.
\cite{eschrigh1,efetov1,giaz1,giaz2,giaz3,alidoust3} The singlet
Cooper pair amplitudes oscillate and simultaneously decay in the
vicinity of the ferromagnet ($F$)-superconductor ($S$) 
interface.\cite{demler,halt} This decaying oscillatory behavior
leads to interesting and intriguing phenomena such as $0$-$\pi$
transitions which can take place by varying the system temperature,
Thouless energy, exchange field, degree of magnetization
inhomogeneity, or inelastic impurities.
\cite{buzdin1,buzdin2,bergeret1,golubov1,eschrigh1,ryaz0,Birge,nazarov}
These $\pi$-junctions have shown promise as building blocks for
quantum computing,\cite{makhlin} thus resulting in extensive studies
of these systems in the clean,  diffusive, and nonequilibrium regimes.
\cite{Bobkov,brataas1,Barash,Cottet,Crouzy,Fominov,Radovic1,Sellier,Pugach2,Jin,eschrigh3}

In a uniform $F$ layer that is proximity coupled to a singlet
superconductor, the pair wavefunction is composed of
an
odd-frequency triplet component in
addition to the usual even-frequency singlet component.\cite{bergeret1,bergeret2,Asano1,Asano2}
The only triplet correlations that can exist in this case are those
with zero total
spin projection $m = 0$ on the spin quantization axis.
Both the  singlet superconducting
correlations and this type of odd-frequency triplet correlations oscillate and sharply decay
inside the $F$ layer.
\cite{bergeret1,bergeret2,halterman1,efetov1} However, if the
magnetization of an $F$ layer possesses an inhomogeneous texture, another 
triplet component can arise which has 
non-zero ($m=\pm 1$) spin projection along the spin 
quantization
axis.\cite{bergeret1,efetov1,Lofwander1,Lofwander2,Kontos,Sosnin}
These triplet correlations are shown to 
penetrate deep into a diffusive $F$ medium with a penetration length 
the same as conventional singlet correlations
in a normal metal.\cite{rob5,Asano2}

The existence
of such triplet correlations have also been observed in experiments, including 
the
measurement of a triplet supercurrent flowing through Holmium hybrid 
structures.\cite{rob1,rob2,rob3,rob4,rob5}  
Shortly thereafter, theoretical works explained these findings 
\cite{rob2,rob3,rob4} in terms of spin triplet proximity effects, 
extending previous studies involving 
inhomogeneous magnetization patterns \cite{rob5,alidoust1}. Triplet
correlations can also be generated in half-metallic systems due to
spin-active
interfaces.\cite{Keizer,Lofwander1,halterman2,brataas1,eschrigh3}
Recently it has been predicted theoretically that these types of
triplet correlations can arise in ballistic bilayers of ferromagnets
with different thicknesses attached to $s$-wave
superconductors.\cite{Trifunovic,Trifunovic2,Hikino,Houzet2} Such
spin superconducting correlations are therefore of interest because they might
play important an important role in dissipationless spintronic
devices.\cite{Hikino,eschrigh1,efetov1,alidoust3}

Recently, a new 
class of Josephson junctions have been experimentally realized in
systems consisting of an $IsF$ section ($I$: insulator layer, $sF$:
a stacked layer that shows superconducting ($s$) and ferromagnetic
properties) sandwiched between two $S$
terminals.\cite{ryaz1,ryaz2,ryaz3,ryaz4} It was shown that
the system can operate as a series of $SIs$ and $sFS$ junctions whose 
properties can be controlled by the thickness of middle $s$ layer.
This type of system was also recently studied
theoretically,\cite{Kupriyanov1} and two operating modes 
were found depending on the critical thickness of the $s$ layer,
equal to $\pi\xi_S/2$, where $\xi_S$ is the superconducting
coherence length. Also motivated by the
experiments  above,
a theoretical work investigated
the
tunability of the magnetic moment
due to the
triplet correlations by varying the superconducting phase
difference of the
outer $S$ banks in \textit{symmetric} layered $SFSFS$, and $SFFSFFS$ structures.\cite{pugach1} 

\begin{figure}[t!]
\includegraphics[width=7.5cm,height=5.50cm]{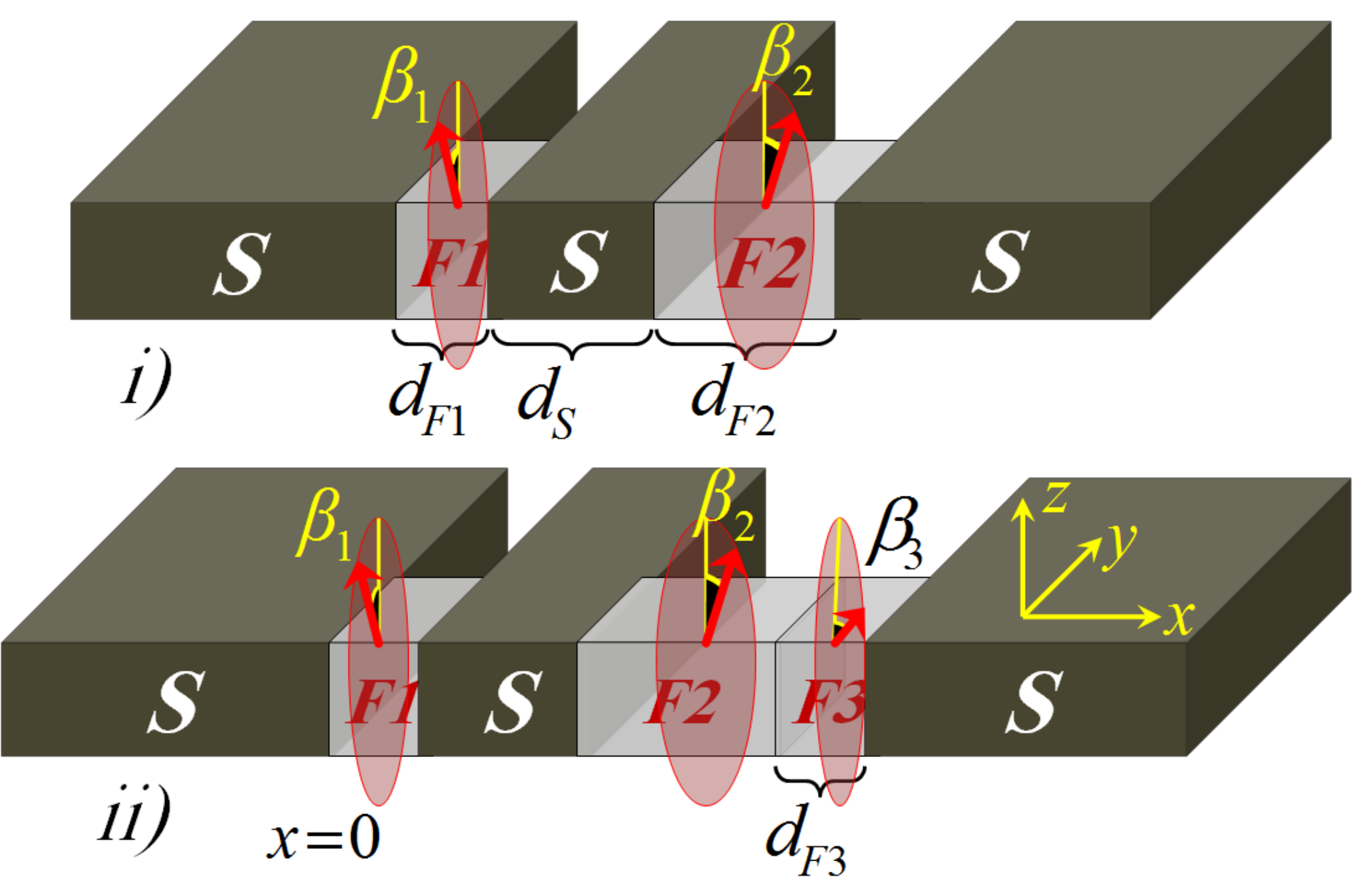}
\caption{\label{fig:model} (Color online) Schematic of the two types
of  Josephson junctions considered: \textit{i)} $SFSFS$ and
\textit{ii)} $SFSFFS$. We assume the $SF$ interfaces are located in
the $zy$ plane while the $x$ axis is along the direction of
Josephson current flow. The length of the middle superconductor $S$
and ferromagnets $F$ are $d_S$, $d_{F1}$, $d_{F2}$, and $d_{F3}$, 
respectively.
Throughout the paper, we denote the  ferromagnetic regions
by $F_1$, $F_2$, and $F_3$ as labeled.
The exchange field of the magnetic layers is assumed
to have arbitrary orientation,
$\vec{h}=(h_x,h_y,h_z)=h_0(\cos\gamma\sin\beta,\sin\gamma\sin\beta,\cos\beta)$,
in which $h_0$ is the amplitude of the exchange field. To analyze the
system properties without loss of generality, all magnetizations are
considered to reside in the $zy$ plane where $\gamma=0$, and
consequently the magnetization orientations can be described solely by 
$\beta$. We therefore define $\beta_{1,2,3}$ for each magnetic
layer.}
\end{figure}

If the superconductivity in
the middle $S$ layer of a $SFSFS$ nanojunction 
is not externally controlled, a self-consistent approach\cite{kh} is
needed to properly determine the magnitude and phase of the
superconducting pair correlations \cite{Kupriyanov1}.
This situation can be realized 
by constructing a stack of three layers ($FsF$) where the middle $s$
layer exhibits
superconducting properties below a critical temperature while the
other layers are insensitive to temperature.
Therefore, by
sandwiching the $FsF$ sample between two $S$ banks and cooling the
system temperature below the critical temperature, proximity induced
modifications arise in the central
layer.\cite{ryaz1,ryaz2,ryaz3,ryaz4}
This 
class of
configurations and approach used
is in contrast to a setup   
where the macroscopic phase in the middle $S$ layer is assumed to be
controlled externally. \cite{pugach1} 
Three-terminal Josephson junctions have been experimentally
realized in the search for Majorana Fermions,\cite{arxiv_snsns}
in
Superconductor/Semiconductor heterostructures.
\cite{major1,major2,major3,major4}
In this work, we also assume  that
supercurrents are generated via  three external
superconducting
terminals.
This is clarified in
Fig.~\ref{fig:model},
where we illustrate our setup  
for $F$ layers that are
sandwiched between the superconducting leads. 
We moreover assume that the system has \textit{no symmetry} 
in configuration
space along the $x$-axis, thus requiring full 
numerical methods to
precisely determine the supercurrent
transport characteristics.  

We consider both $SFSFS$ and $SFSFFS$ type junctions in 
the diffusive regime. We demonstrate that the transport of
supercurrent in each $F$ region can be easily controlled by the
relative magnetization orientation of one of the $F$ layers. We show
that this valve effect follows in part from the 
triplet components involved in supercurrent
transport
arising from
the superconducting phase gradients present among the $S$
terminals.
Throughout our calculations we have assumed that the
macroscopic phase of the three superconducting terminals can be
externally varied, and hence the charge current is not necessarily
conserved within
the $S$ regions.
In the $F$ regions, the charge current is constant, but the
spin-current is in general not conserved due to
the exchange interaction.
We employ the 
Keldysh-Usadel quasiclassical method in the diffusive limit to study
these multilayer systems.
We then decompose the total supercurrent into both its even- and 
odd-frequency components, and investigate their spatial 
profiles as a function of various values of magnetization
orientations and phase differences. We demonstrate that  the 
total charge supercurrent in one $F$ can change sign by means of
magnetization rotation in one of the other $F$ layers, while the total
charge supercurrent does not undergo a reversal in the 
rotated
$F$ layers (or vice a versa). 
This behavior of the current indicates that it is possible to
arrange a sequence of controllable $0$ and $\pi$ Josephson junctions
in a three terminal $SFSFS$ spin switch. By studying the current
components as a function of position, we are able to pinpoint the
origin of the spin-controlled supercurrent. Typically in the middle
$S$ region, the singlet contribution to the supercurrent follows a
nearly linear spatial variation, while
the nonvanishing odd-frequency triplet components 
do not decay in space.

We are  able to extract from our numerical results analytical
expressions for the current-phase relations, thus
simplifying the overall physical picture. 
The numerical solutions
showed
that all components of
the supercurrent
are described by a simple sinusoidal relation that
depends on the
differences of the phases, $\theta_L,\theta_R$, and
$\theta_M$, corresponding to the
left, right, and middle $S$ regions respectively 
(see Fig. \ref{fig:model}).  
We found that an additional term arises in the current-phase relations besides the
regular sinusoidal terms,
which for sufficiently
thick middle $S$
electrodes is responsible for spin-controlled transport through the
junction.
Although this additional term is present for all $d_S$, its signatures
are more prominent when quasiparticle
tunneling between outer $S$ electrodes is suppressed,
corresponding to the regime of
large $d_S$. Therefore, depending on the
superconducting phase differences involving $\theta_L,\theta_R$, and
$\theta_M$, a 
relatively thick middle $S$ electrode
can limit the spin-controlled features described above. 

The paper is organized as follows. In Sec. \ref{section:theory} we
discuss the method employed, details of our assumptions, and
technical points used in our calculations. In Sec. \ref{results} we
discuss our  results, analyze them and suggest possible 
applications of our findings. We finally summarize and give
the concluding remarks in Sec. \ref{conclusion}.

\section{Theory and Methods}\label{section:theory}

In this section, we outline the assumptions present and the
theoretical approach used to study $SFSFS$ and $SFSFFS$ type
systems. The Keldysh-Usadel technique employs the total Green's
function with three blocks labeled Retarded ($R$), Advanced ($A$),
and Keldysh ($K$). Using the labeled blocks, the total Green's
function is represented by\cite{bergeret1};
\begin{equation}\label{eq:greenfunc}
    \hat{G}(\vec{r},\varepsilon,T)=\left(\begin{array}{cc}
              G^{A} & G^{K} \\
              \mathbf{0} & G^{R}
            \end{array}\right),\;G^{R}=\left(\begin{array}{cc}
              g^{R} & f^{R} \\
              -\tilde{f}^{R} & -\tilde{g}^{R}
            \end{array}\right).
\end{equation}
The propagators are position, $\vec{r}$, and temperature, $T$, 
dependent. The quasiparticles' energy is denoted by $\varepsilon$
and is measured from Fermi level. In the equilibrium steady state,
the advanced and Keldysh blocks can be related via
$G^{A}=-(\hat{\rho}_3G^R\hat{\rho}_3)^{\dag}$ and
$G^{K}=\tanh(\beta\varepsilon)(G^{R}-G^{A})$ in which $\hat{\rho}_3$
is the third component of Pauli matrices
$\vec{\hat{\rho}}=(\hat{\rho}_1,\hat{\rho}_2,\hat{\rho}_3)$ (see
Appendix) and $\beta=k_BT/2$, with $k_B$ the Boltzmann constant. In
the absence of a ferromagnetic exchange field, the total Green's
function reduces to a $4\times 4$
propagator.\cite{bergeret1,efetov1} However, in the presence of a
general exchange field term, the total Green's function becomes a
$8\times 8$ matrix.\cite{rob2} In the regime in which proximity
effects are small, we may
expand the Green's function around the bulk solution\cite{bergeret1} 
$\hat{G}_0=\text{diag}(1,-1)$, i.e. $\hat{G}\approx
\hat{G}_0+\hat{f}$. In this approximation we arrive at;
\cite{Lofwander2,linder1,alidoust4}
\begin{align}\label{Advanced Gree}
\hat{G}^{A}=\begin{pmatrix}
-1 & 0 & -f^{R}_{\upuparrows}(-\varepsilon) & f^{R}_{-}(-\varepsilon) \\
0 & -1  &f^{R}_{+}(-\varepsilon)  & -f^{R}_{\downdownarrows}(-\varepsilon)  \\
 [f^{R}_{\upuparrows}(\varepsilon)]^{\ast}   & -[f^{R}_{-}(\varepsilon)]^{\ast}   & 1  &  0  \\
-[f^{R}_{+}(\varepsilon)]^{\ast}& [f^{R}_{\downdownarrows}(\varepsilon)]^{\ast}  & 0  &  1  \\
\end{pmatrix},
\end{align}
in which the asterisk denotes complex conjugation. The arrays with
$\upuparrows$, and $\downdownarrows$ correspond to the spin-one
(equal-spin) components while those with $\pm$ represent the
superconducting correlations with zero spin (opposite-spin
pairing).\cite{Lofwander2,efetov2}

The general form of Usadel equation \cite{Usadel} (which can be
derived from the Eilenberger equation \cite{Eilenberger}) in the
presence of an exchange field with components
$\vec{h}=(h_x,h_y,h_z)$ in the ferromagnetic layers, and a gap
energy $\Delta$ associated with the $s$-wave superconducting region,
can be compactly expressed by\cite{morten}, 
\begin{align}\label{eq:usadel}
D[\hat{\partial},\hat{G}[\hat{\partial},\hat{G}]]+i[ \varepsilon
\hat{\rho}_{3}-\hat{\Delta}+
\text{diag}[\vec{h}\cdot\vec{\tau},(\vec{h}\cdot\vec{\tau})^{T}],\hat{G}]=0,
\end{align}
where $T$ denotes transpose, $\hat{\rho}_{3}$ and $\vec{\tau}$ are $4\times 4$ and $2\times
2$ Pauli matrices, respectively.
The matrices are defined in the Nambu and spin spaces which are 
given in Appendix. Here $D$ is diffusive constant of the highly
impure medium and the brackets imply commuter algebra.\cite{morten}
The gradient operator is written shorthand as $\hat{\partial}$, such
that $\hat{\partial}=(\partial_x,\partial_y,\partial_z)$, which for
our one dimensional system reduces simply to $\partial_x$. Here
$\hat{\Delta}$ is a $4\times 4$ matrix that is defined as
follows:\cite{morten}
\begin{equation}
\hat{\Delta}=\left(\begin{array}{cccc}
                     0 & 0 & 0 & \Delta \\
                     0 & 0 & -\Delta & 0 \\
                     0 & \Delta^\ast & 0 & 0 \\
                     -\Delta^\ast & 0 & 0 & 0
                   \end{array}
 \right).
\end{equation}
In the ferromagnet regions, the superconducting gap energy $\Delta$
in Eq. (\ref{eq:usadel}), should be equal to zero while in the
diffusive nonmagnetic superconducting layers, the exchange energy
$\vec{h}$ is set equal to zero. The proximity effect that governs
the interaction between the differing media is accounted for by the
appropriate boundary conditions at the junctions and interfaces.  To
accurately model realistic barrier regions,  
we use  Kupriyanov-Lukichev boundary conditions at both $SF$
interfaces near the end of the sample;\cite{cite:zaitsev}
\begin{equation}\label{eq:bc}
    \zeta(\hat{G}\hat{\partial}\hat{G})\cdot\vec{n}=[\hat{G}_{\text{BCS}}(\theta_{L,R}),\hat{G}],
\end{equation}
where $\vec{n}$ is a unit vector normal to the interface. The leakage of
correlations are governed by the parameter $\zeta$, which depends on
the resistance of the interface and the diffusive normal
region.\cite{linder1,alidoust3,alidoust4} The bulk solution,
$\hat{G}_{\text{BCS}}(\theta_{L,R})$, for an $s$-wave superconductor
is;\cite{morten}
\begin{eqnarray} \label{Eq:bulk_solution}
&&\hat{G}^{R}_{\text{BCS}}(\theta_{L,R})=\nonumber\\&&\left[
                  \begin{array}{cc}
                   \mathbf{1}\cosh(\vartheta_{L,R}(\varepsilon)) & i\tau_2\sinh(\vartheta_{L,R}(\varepsilon))e^{i\theta_{L,R}} \\
                    i\tau_2\sinh(\vartheta_{L,R}(\varepsilon))e^{-i\theta_{L,R}} & -\mathbf{1}\cosh(\vartheta_{L,R}(\varepsilon)) \\
                     \end{array}
                                \right],
\end{eqnarray}
\begin{equation}
\nonumber
\vartheta_{L,R}(\varepsilon)=\text{arctanh}(\frac{\mid\Delta_{L,R}\mid}{\varepsilon}).
\end{equation}
We write $\Delta_{L,R}$ for the superconducting gap in the leftmost ($L$) and
rightmost ($R$) bulk superconductors. On the other hand, we assume that the
other interfaces are fully transparent (no insulating layer) 
for both composite $SFS$ Josephson junction configurations.

The Usadel equation in the general form given above, involving the
magnetic exchange field with arbitrary orientation, leads to $8$
coupled complex partial differential equations, even in the
low proximity limit where the equations can be linearized. It should
be reiterated in passing that the interaction between inhomogeneous
ferromagnets and $s$-wave superconductors leads to triplet
correlations with nonzero projection along the spin quantization
axis.\cite{bergeret1} Therefore, we may assume that the Green's
function describing such systems can be considered as a summation of
singlet ($\mathbb{S}$) and triplet ($\vec{\mathbb{T}}$) components
(spin parameterization).\cite{bergeret1,Lofwander2} We thus write
\cite{Lofwander2,efetov2}:
\begin{eqnarray} \label{decomp}
\hat{f}(\varepsilon)=
i(\mathbb{S}(\varepsilon)+\vec{\mathbb{T}}(\varepsilon).\vec{\tau})\tau_y,
\end{eqnarray}
where $\vec{\tau}=(\tau_x,\tau_y,\tau_z)$ is a vector comprised
of Pauli
matrices. If we now substitute this decomposition of the
anomalous
Green's function into the Usadel equation Eq. (\ref{eq:usadel}), we
end up with the following coupled set of differential equations:
\begin{widetext}
\begin{eqnarray}\label{Linearized Usadel Eq.}
&&D\left\{\mp\partial_x^{2} \mathbb{T}_{x}(-\varepsilon)+i\partial_x^{2} \mathbb{T}_{y}(-\varepsilon)\right\}+i\left\{-2 \varepsilon (\mp \mathbb{T}_{x}(-\varepsilon)+i \mathbb{T}_{y}(-\varepsilon)) \mp 2\mathbb{S}(-\varepsilon) (h_{x} \mp i h_{y}) \right\}=0,\\
&&D\left\{\mp\partial_x^2\mathbb{S}(-\varepsilon)+\partial_x^2\mathbb{T}_z(-\varepsilon)\right\}+i\left\{\mp2\mathbb{T}_x(-\varepsilon) h_{x} \mp 2 \mathbb{T}_{y}(-\varepsilon) h_{y} - 2 (\mp\mathbb{S}(-\varepsilon)+\mathbb{T}_z(-\varepsilon)) (\varepsilon \pm h_{z}) \right\}=\pm 2i\Delta_Me^{i\theta_M},\\
&&D\left\{\mp\partial_x^{2}\mathbb{T}_{x}^{\ast}(\varepsilon)-i\partial_x^{2}\mathbb{T}_{y}^{\ast}(\varepsilon)\right\}+i\left\{\pm 2 (h_{x} \pm i h_{y}) \mathbb{S}^{\ast}(\varepsilon) - 2 \varepsilon (\mp\mathbb{T}_{x}^{\ast}(\varepsilon)-i\mathbb{T}_{y}^{\ast}(\varepsilon)) \right\}=0,\\
&&D\left\{\mp\partial_x^{2}\mathbb{S}^{\ast}(\varepsilon)+\partial_x^2\mathbb{T}_{z}^{\ast}(\varepsilon)\right\}+i\left\{
2 (-\varepsilon \pm
h_{z})(\mp\mathbb{S}^{\ast}(\varepsilon)+\mathbb{T}_{z}^{\ast}(\varepsilon))
\pm 2 h_{x} \mathbb{T}_{x}^{\ast}(\varepsilon) \pm 2 h_{y}
\mathbb{T}_{y}^{\ast}(\varepsilon)\right\}=\mp 2i \Delta_M
e^{-i\theta_M}.
\end{eqnarray}
\end{widetext}
Since we need to solve the Usadel equations in the central $S$
layer, we denote the superconducting gap in this region by
$\Delta_M$, with macroscopic phase $\theta_M$. If the decomposition
in Eq.~(\ref{decomp}) is substituted into the Kupriyanov-Lukichev
boundary conditions (Eq. (\ref{eq:bc})), the following differential
equations must be satisfied at the left $SF$
interface:\cite{alidoust4}
\begin{eqnarray}\label{bc_F1}
&&(\zeta\partial_x  - c^{\ast}(\varepsilon))(\mp \mathbb{T}_{x}(-\varepsilon)+i \mathbb{T}_{y}(-\varepsilon))=0,\\
&&(\zeta\partial_x  - c^{\ast}(\varepsilon))(\mp\mathbb{S}(-\varepsilon)+\mathbb{T}_z(-\varepsilon)) =\mp s^{\ast}(\varepsilon), \\
&&(\zeta\partial_x  - c^{\ast}(\varepsilon))(\mp\mathbb{T}_{x}^{\ast}(\varepsilon)-i\mathbb{T}_{y}^{\ast}(\varepsilon))=0,\\
&&(\zeta\partial_x  -
c^{\ast}(\varepsilon))(\mp\mathbb{S}^{\ast}(\varepsilon)+\mathbb{T}_{z}^{\ast}(\varepsilon))
=\pm s^{\ast}(\varepsilon).
\end{eqnarray}
Here we define the following expressions for $s(\varepsilon)$ and
$c(\varepsilon)$,\cite{bergeret1,buzdin1,morten}
\begin{eqnarray}
&& s(\varepsilon)=\frac{-\Delta\text{sgn}(\varepsilon)\Theta(\varepsilon^2-\Delta^2)}{\sqrt{\varepsilon^2-\Delta^2}}+\frac{i\Delta\Theta(\Delta^2-\varepsilon^2)}{\sqrt{\Delta^2-\varepsilon^2}},\\
&&
c(\varepsilon)=\frac{\mid\varepsilon\mid\Theta(\varepsilon^2-\Delta^2)}{\sqrt{\varepsilon^2-\Delta^2}}-\frac{i\varepsilon\Theta(\Delta^2-\varepsilon^2)}{\sqrt{\Delta^2-\varepsilon^2}},
\end{eqnarray}
in which $\Theta(x)$ represents a step-function. Likewise,
performing the same decomposition for the right $FS$ interface and
assuming that both outer superconducting terminals have equal
superconducting gaps $\Delta_L=\Delta_R=\Delta$, the
Kupriyanov-Lukichev boundary conditions become:\cite{alidoust4}
\begin{eqnarray}\label{bc_F2}
&&(\zeta\partial_x  + c^{\ast}(\varepsilon))(\mp \mathbb{T}_{x}(-\varepsilon)+i \mathbb{T}_{y}(-\varepsilon))=0,\\
&&(\zeta\partial_x  + c^{\ast}(\varepsilon))(\mp\mathbb{S}(-\varepsilon)+\mathbb{T}_z(-\varepsilon)) =\pm s^{\ast}(\varepsilon), \\
&&(\zeta\partial_x  + c^{\ast}(\varepsilon))(\mp\mathbb{T}_{x}^{\ast}(\varepsilon)-i\mathbb{T}_{y}^{\ast}(\varepsilon))=0,\\
&&(\zeta\partial_x  +
c^{\ast}(\varepsilon))(\mp\mathbb{S}^{\ast}(\varepsilon)+\mathbb{T}_{z}^{\ast}(\varepsilon))
=\mp s^{\ast}(\varepsilon) .
\end{eqnarray}
To investigate the system, the transformed coupled differential equations 
and associated boundary conditions must be solved using 
geometrical and material parameters that are  experimentally
appropriate. Unfortunately, this complicated system of coupled
differential equations can be simplified and decoupled for only a
limited range of parameters and configurations. When such
simplifications are possible, the equations have the advantage that
sometimes they can lead to analytical results. However, for our
complicated multilayer configurations, numerical methods are the
most efficient and sometimes the only possible routes to investigate 
the relevant transport properties.

One of the most important physical quantities related to transport
is the supercurrent that is generated from the macroscopic phase
differences between superconducting terminals separated by a ferromagnet. 
\begin{figure*}[t!]
\includegraphics[width=16cm,height=6.0cm]{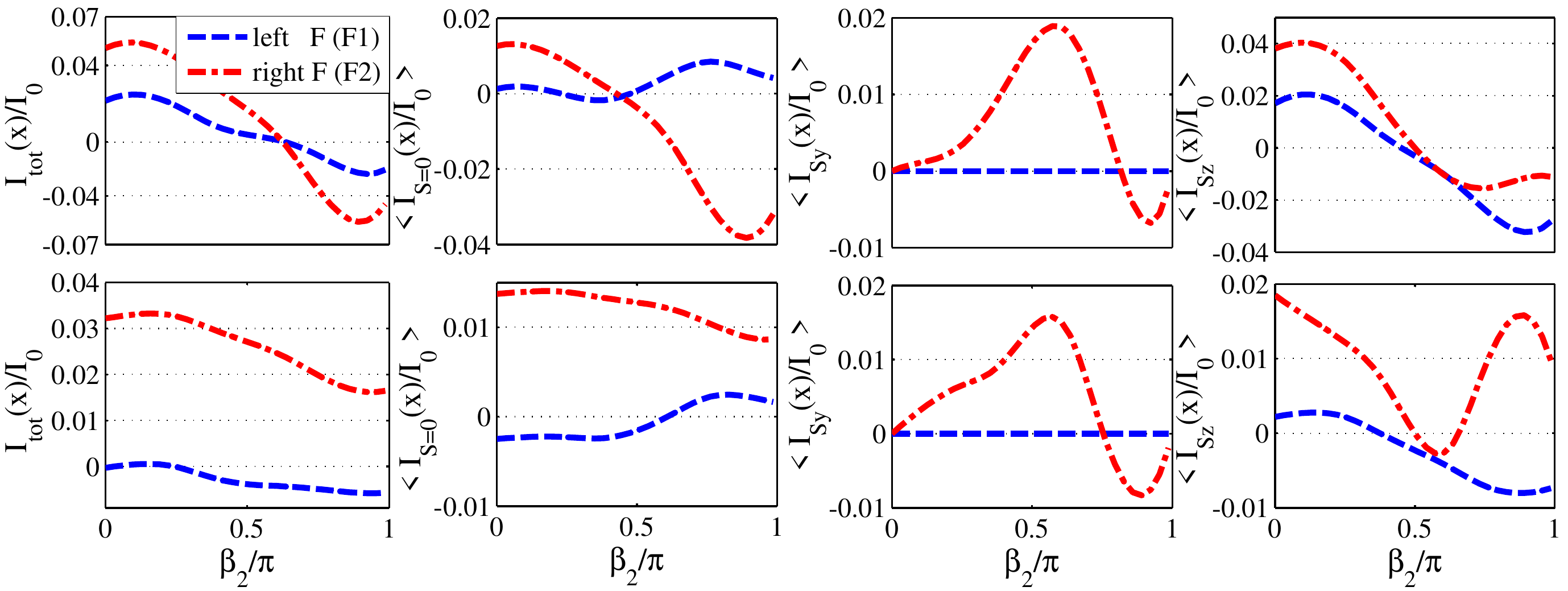}
\caption{\label{fig:sfsfs_beta} (Color online) Total critical
supercurrent
($I_{tot}$),  singlet  ($I_{S0}$), and odd-frequency ($I_{Sy}$, 
$I_{Sz}$) components of the Josephson $SFSFS$ structure. The current
and the average of its components are shown as a function of
exchange field orientation, $\beta_2$  (see Fig.~\ref{fig:model}). 
The averages are taken over the labeled magnetic
regions. In the top row we assume $d_{F1}=d_{F2}=0.4\xi_S$, and 
$d_S=0.2\xi_S$. While in the bottom row $d_{F1}=0.4\xi_S$,
$d_{F2}=0.7\xi_S$, and the thickness of middle superconducting lead
is kept unchanged. We set the superconducting phase of the left, 
and middle superconductors to be $\theta_L=\theta_M=0$ and
$\theta_R=\pi/2$ (corresponding to the maximum supercurrent 
in this case, 
see text), respectively.}
\end{figure*}
To determine the charge supercurrent, we consider the general
expression for the charge current density in the steady state. This
involves the Keldysh component of total Green's function
via,\cite{morten,bergeret1,efetov2}
\begin{equation}\label{eq:currentdensity}
J_x\text{(}x\text{)}=J_{0}\int
d\varepsilon\text{Tr}\{\hat{\rho}_{3}(\hat{G}[\hat{\partial},\hat{G}])^{K}\}.
\end{equation}
Here $J_{0}$ is a normalization constant equal to $e N_0 D/8$ in
which $e$ is the electron charge and $N_0$ is the density of states of 
a normal metal at the Fermi surface. To derive a tractable expression 
for the charge supercurrent density, the Advanced and Keldysh blocks
of the Green's function are obtained from the previously mentioned
relations above that relate the $A$, $R$, and $K$ blocks stemming
from Eq.~(\ref{decomp}). We assume that the current is flowing along
the $x$ axis, normal to the interfaces located in the $yz$ plane
(see Fig.~\ref{fig:model}). After some lengthy calculations, we
arrive at the current density:
\begin{eqnarray}\label{eq:current}
&&\nonumber J(x) =J_0 \int_{-\infty}^{\infty} d\varepsilon
\left\{\right.\\\nonumber && \left.
\mathbb{S}(\varepsilon)\partial_x\mathbb{S}^{\ast}(-\varepsilon) -
\mathbb{S}(-\varepsilon)\partial_x\mathbb{S}^{\ast}(\varepsilon)+
\mathbb{S}(\varepsilon)^{\ast}\partial_x\mathbb{S}(-\varepsilon)
-\right.\\\nonumber && \left.
\mathbb{S}(-\varepsilon)^{\ast}\partial_x\mathbb{S}(\varepsilon)
-\partial_x\mathbb{T}_x(-\varepsilon)\mathbb{T}_x^{\ast}(\varepsilon)
+\partial_x\mathbb{T}_x(\varepsilon)\mathbb{T}_x^{\ast}(-\varepsilon)
-\right.\\\nonumber && \left.
\partial_x\mathbb{T}_x^{\ast}(-\varepsilon)\mathbb{T}_x(\varepsilon)
+\partial_x\mathbb{T}_x^{\ast}(\varepsilon)\mathbb{T}_x(-\varepsilon)
-\partial_x\mathbb{T}_y(-\varepsilon)\mathbb{T}_y^{\ast}(\varepsilon)
+\right.\\\nonumber &&
\left.
\partial_x\mathbb{T}_y(\varepsilon)\mathbb{T}_y^{\ast}(-\varepsilon)
-\partial_x\mathbb{T}_y^{\ast}(-\varepsilon)\mathbb{T}_y(\varepsilon)
+\partial_x\mathbb{T}_y^{\ast}(\varepsilon)\mathbb{T}_y(-\varepsilon)-\right.\\\nonumber
&& \left.
\partial_x\mathbb{T}_z(-\varepsilon)\mathbb{T}_z^{\ast}(\varepsilon)
+\partial_x\mathbb{T}_z(\varepsilon)\mathbb{T}_z^{\ast}(-\varepsilon)
-
\partial_x\mathbb{T}_z^{\ast}(-\varepsilon)\mathbb{T}_z(\varepsilon)+
\right.\\ &&
\left.\partial_x\mathbb{T}_z^{\ast}(\varepsilon)\mathbb{T}_z(-\varepsilon)
\right\}\tanh(\varepsilon\beta/2).
\end{eqnarray}
As can be seen, the integration covers the entire quasiparticle
energy spectrum. To obtain the charge supercurrent, it is necessary
to integrate the current density along $y$ over the junction width
$W$. Since we assume that our system is translationally invariant
along the $y$ direction, the current density must of course also be
$y$-independent. It is convenient then in the results that follow,
to normalize the supercurrent, $I(x)$, by $I_0 \equiv W J_0$. Having
now outlined our general method and numerical approach, we proceed
to present our numerical results and study the supercurrent for
particular cases of $SFSFS$ and $SFSFFS$ Josephson junctions.

\section{Results}\label{results}

In presenting our numerical results,
we
decompose the general charge
supercurrent density [Eq.~(\ref{eq:current})]
into each of its four components.
The associated supercurrent subsequently has the components,
\begin{eqnarray}\label{eq:components}
&&\nonumber I_{S0}(x) =I_0 \int_{-\infty}^{\infty} d\varepsilon
\left\{
\mathbb{S}(\varepsilon)\partial_x\mathbb{S}^{\ast}(-\varepsilon) -
\mathbb{S}(-\varepsilon)\partial_x\mathbb{S}^{\ast}(\varepsilon)+\right.\\
&& \left.
\mathbb{S}^{\ast}(\varepsilon)\partial_x\mathbb{S}(-\varepsilon)
-\mathbb{S}^{\ast}(-\varepsilon)\partial_x\mathbb{S}(\varepsilon)\right\}\tanh(\varepsilon\beta/2),\\\nonumber 
&&\nonumber I_{Sx}(x) =I_0 \int_{-\infty}^{\infty} d\varepsilon
\left\{
-\partial_x\mathbb{T}_x(-\varepsilon)\mathbb{T}_x^{\ast}(\varepsilon)
+\partial_x\mathbb{T}_x(\varepsilon)\mathbb{T}_x^{\ast}(-\varepsilon)
\right.\\ && \left.
-\partial_x\mathbb{T}_x^{\ast}(-\varepsilon)\mathbb{T}_x(\varepsilon)
+\partial_x\mathbb{T}_x^{\ast}(\varepsilon)\mathbb{T}_x(-\varepsilon)\right\}\tanh(\varepsilon\beta/2),\\
&&\nonumber I_{Sy}(x) =I_0 \int_{-\infty}^{\infty} d\varepsilon
\left\{-\partial_x\mathbb{T}_y(-\varepsilon)\mathbb{T}_y^{\ast}(\varepsilon)
+
\partial_x\mathbb{T}_y(\varepsilon)\mathbb{T}_y^{\ast}(-\varepsilon)\right.\\ && \left.
-\partial_x\mathbb{T}_y^{\ast}(-\varepsilon)\mathbb{T}_y(\varepsilon)
+\partial_x\mathbb{T}_y^{\ast}(\varepsilon)\mathbb{T}_y(-\varepsilon)\right\}\tanh(\varepsilon\beta/2),\\
&&\nonumber I_{Sz}(x) =I_0 \int_{-\infty}^{\infty} d\varepsilon
\left\{\partial_x\mathbb{T}_z(-\varepsilon)\mathbb{T}_z^{\ast}(\varepsilon) 
+\partial_x\mathbb{T}_z(\varepsilon)\mathbb{T}_z^{\ast}(-\varepsilon)
\right.\\ &&
\left.-\partial_x\mathbb{T}_z^{\ast}(-\varepsilon)\mathbb{T}_z(\varepsilon)+
\partial_x\mathbb{T}_z^{\ast}(\varepsilon)\mathbb{T}_z(-\varepsilon)
\right\}\tanh(\varepsilon\beta/2).
\end{eqnarray}
The total charge current, $I_{tot}(x)$, is thus the sum,
\begin{equation}
I_{tot}(x)=I_{S0}(x)+I_{Sx}(x)+I_{Sy}(x)+I_{Sz}(x),
\end{equation}
where $I_{S0}$ denotes the singlet supercurrent component. We take the axis of spin quantization 
to lie along the $z$ direction throughout the whole system, and thus
the components $I_{Sx}$, $I_{Sy}$, represent the equal-spin triplet
components with total spin projection of $m=\pm1$ on the axis of
spin quantization, while $I_{Sz}$, corresponds to opposite spin
triplets with $m=0$, and a total spin projection of
zero.\cite{bergeret1,eschrigh3,efetov1,efetov2,linder1}
The decomposition of the supercurrent 
into the singlet and triplet components
can also serve to identify the long range contributions to the
supercurrent.\cite{eschrigh3,efetov2,linder1,alidoust4} 

To begin, we first consider
the simpler $SFSFS$ junction (Fig.~\ref{fig:model}, part $i)$).
We assume the far left $SF$ interface is located at $x=0$ and
all interfaces reside in the $yz$ plane. The thickness of $F_1$, $F_2$, and
the middle superconducting lead are
denoted by $d_{F1}$, $d_{F2}$, and $d_S$, respectively. Our
theoretical framework permits each $F$ layer to possess a general
exchange field with arbitrary orientation,
$\vec{h}_{1,2}=h_0(\cos\gamma_{1,2}\sin\beta_{1,2},\sin\gamma_{1,2}\sin\beta_{1,2},\cos\beta_{1,2})$.
To study concrete examples, we consider systems with in-plane
magnetization orientations
where $\gamma_{1,2}=0$, and thus rotation occurs in the $yz$ plane.
This also implies
that $\beta_1$ and $\beta_2$ fully characterize the magnetization
orientations of $F_1$ and $F_2$, respectively, as illustrated in
Fig.~\ref{fig:model}. The magnetization
orientation of $F_1$ is assumed fixed in the $z$ direction
($\beta_1=0$), while the magnetization in $F_2$ rotates with angle
$\beta_2$. We assume that the proximity effects related 
to the flow of 
spin-polarized supercurrent into the ferromagnetic regions has a
negligible effect on their respective magnetizations.\cite{pugach1}
This assumption is frequently used in most of the theoretical works
on the ferromagnetic multilayer Josephson
configurations.\cite{buzdin1,bergeret1} As mentioned above, we also
assume the macroscopic phases of the three superconducting leads
(left, middle, and right), $\theta_L$, $\theta_M$, $\theta_R$, are
controlled externally.\cite{pugach1} Throughout our calculations, we
consider a low temperature of $T=0.05T_c$, and a ferromagnetic
strength given by the exchange field magnitude
$|\vec{h}_{1,2}|=h_0=10\Delta_0$. Here $T_c$ is the superconducting
critical temperature and $\Delta_0$ is the superconducting gap at
zero temperature, also we set $k_B=\hbar=1$. In this paper, all
energies are normalized by $\Delta_0$ while lengths are normalized
by $\xi_S$, the superconducting coherence length. 
Since we have considered the low proximity  
tunneling limit in the diffusive regime, the interface
transparencies affect the strength of the leakage of superconducting
proximity correlations. In our actual calculations we have set
$\zeta=4$ which is consistent with the low proximity limit.

\begin{figure*}[t]
\includegraphics[width=17.0cm,height=7.60cm]{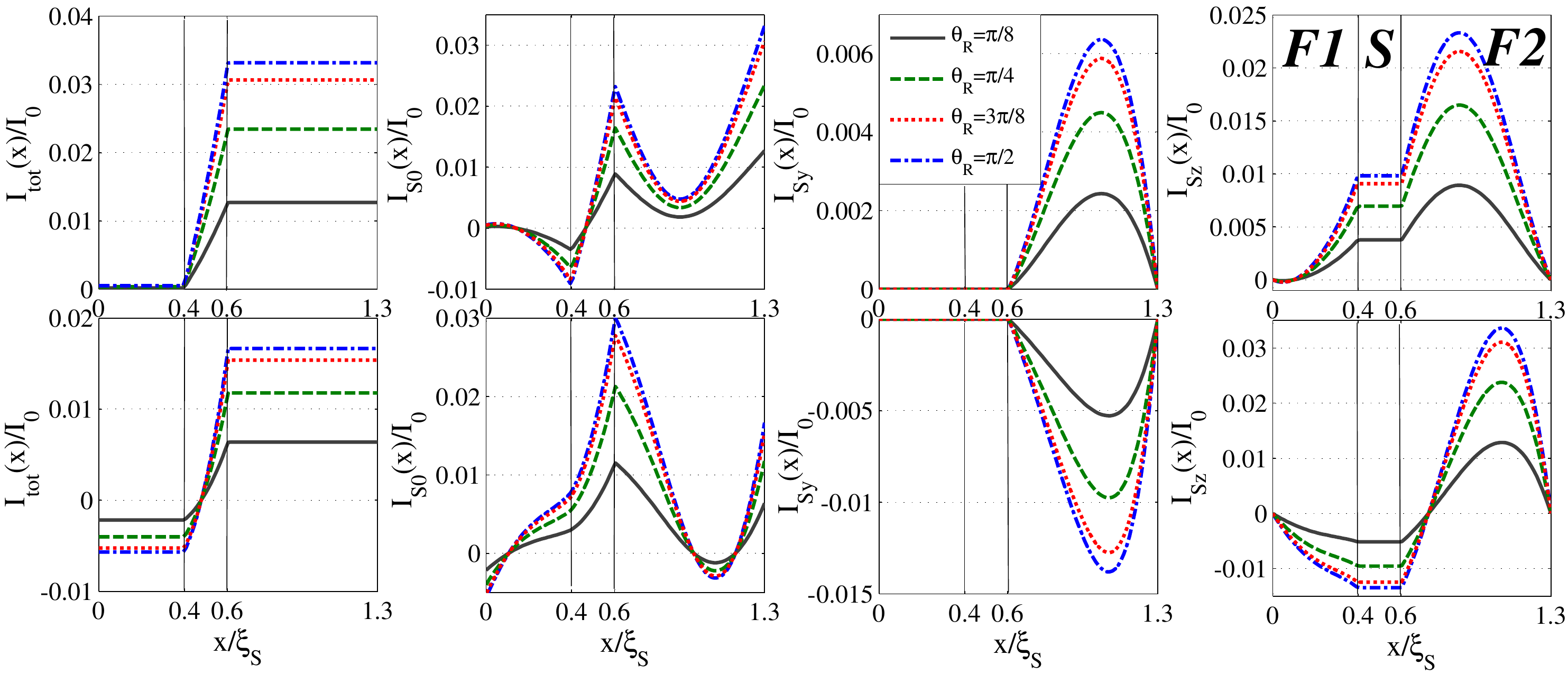}
\caption{\label{fig:sfsfs_x} (Color online) The Josephson current
and its
components (singlet ($I_{S0}$) and odd-frequency triplet ($I_{Sy,z}$)) vs position in 
an $SFSFS$ junction for four values of the superconducting phase in
the right terminal: $\theta_R=$ $\pi/8$, $\pi/4$, $3\pi/8$, $\pi/2$.
We set $d_{F1}=0.4\xi_S$, $d_{F2}=0.7\xi_S$, $d_S=0.2\xi_S$, and
$\theta_L=\theta_M=0$. The top row represents results where 
$\beta_2=\pi/8$ while the bottom row shows results for
$\beta_2=7\pi/8$. Throughout the paper we assume the exchange field
direction in $F_1$ is oriented in the $z$ direction, so that
$\beta_1=0$. Different magnetic and superconducting regions are
separated by vertical lines and marked by $F_1$, $F_2$, and $S$.}
\end{figure*}
We first discuss the current phase relations, which are important
for
determining the experimentally relevant critical current, and 
the fundamental nature of resistanceless transport through
junctions. When permissible, exact analytical current phase
relationships can reveal more about the fundamental physics
involved, and the important role of phase-coherent transport in
Josephson junctions for practical device development.
Also, since we have considered the low proximity limit in the 
diffusive regime, higher order
harmonics are washed out. \cite{buzdin1} 
As mentioned in passing, exact analytical
expressions are generally impossible in the types of systems considered here due to the
complicated complex partial differential equations involved.
Nonetheless, we were 
still able to
extract simple current-phase relations from 
the full numerical
results.
We found that if the thickness of the middle superconductor is
sufficiently thin, the coupling between the two outer
superconductors results in supercurrent flow in the magnetic regions
that obeys
sinusoidal current-phase relations involving combinations of the three 
superconducting phases.  Our numerical 
investigations have found that for our regimes of interest, the current phase relation in 
$F_1$ region obeys;
\begin{equation}\label{eq:sinusoidal}
   I_L=
   I_{1}\sin\varphi_{LR}+I_{2}\sin\varphi_{LM}+I_{3}\sin\varphi_{RM}\sin\varphi_{LM},
   \end{equation}
where $\varphi_{LR}=\theta_L-\theta_R$,
$\varphi_{LM}=\theta_L-\theta_M$, and $\varphi_{RM}=\theta_R-\theta_M$. 
Here $I_{1,2,3}$ are constants which in general depend on geometry
($d_{F1}$, $d_{F2}$, $d_S$), temperature $T$, exchange fields
$\vec{h}_{1,2}$,
and interface transparencies $\zeta$. 
In determining the current-phase relation above, several  systematic
investigations were numerically performed involving the  macroscopic
phases in each of the three terminals.
Of the three phases, $\theta_R$, $\theta_L$, and $\theta_M$,
the supercurrent is calculated by varying one phase, e.g., $\theta_R$,
while the
other two are kept fixed.
This process is repeated for
several differing fixed phases (say, $\theta_L$ and 
$\theta_M$). This procedure also reveals the precise form of the
coefficients $I_{1,2,3}$, which generally vary as the system
parameters change. Below we present concrete examples that
illustrate the relevant terms in the current-phase relation of
Eq.~(\ref{eq:sinusoidal}).
Unless otherwise noted, we set 
$\theta_L=\theta_M=0$ and numerically vary $\theta_R$. 
The precise nature of supercurrent transport in our three-terminal
spin switch
hinges crucially on not only the phase of the central $S$ layer, but also its width. 
The geometrical effects of the central $S$ layer can be seen in the
limit of large $d_S$ ($d_S\gg\xi_S$) where we find 
$I_{L}=0$. This can be understood by noting that since
$\varphi_{LM}=0$, we have $ I_L= I_{1}\sin\varphi_{LR}$.
Consequently for large $d_S$, the tunneling of quasiparticles
through the middle $S$ region is highly suppressed, giving $I_1=0
\Rightarrow I_L=0$. For middle layers that are moderately thick (on
the order of a few $\xi_S$), the last term in
Eq.~(\ref{eq:sinusoidal}) implies (for $\varphi_{LM},\varphi_{RM}
\neq 0$) the two outer superconducting terminals are not entirely
isolated from one another however. The effects of this
coupling-term 
will be discussed in more detail below. Although we consider three
superconducting terminals in serial, our findings are consistent
with Ref.~\onlinecite{alidoust2} where a cross diffusive
ferromagnetic four-terminal Josephson transistor is
studied. We have found that the sinusoidal relations are generally 
valid for 
the supercurrent when the relative magnetizations of the $F$ layers
are non-collinear. The
relations (Eq.~(\ref{eq:sinusoidal})) can thus be considered guides in 
determining  phase differences that  lead to optimal current flow.
One such possibility involves the choice of $\theta_R=\pi/2$, which
according to the sinusoidal relations, corresponds to maximum
supercurrent flow, or equivalently the critical current, for the
case when $\theta_L=\theta_M=0$, 
and where the middle $S$ is sufficiently thin ($I_1\neq 0$, 
see Fig. \ref{fig:sfsfs_beta}). In
other words, the critical current in a moderately thin middle $S$ 
electrode and  fixed $\theta_L=\theta_M=0$ occurs at
$\theta_R=\pi/2$.
We discuss below the benefits of situations 
where $\theta_L\neq\theta_M$.

Figure~\ref{fig:sfsfs_beta} exhibits the total Josephson current and
its spatially averaged components through both ferromagnetic regions
of the $SFSFS$ system (Fig. \ref{fig:model} $i)$), versus 
magnetization orientation of $F_2$ layer. We assume the
magnetization orientation of $F_1$ is fixed
along the $z$ axis (spin quantization axis) while the exchange field
in $F_2$ makes an angle $\beta_2$ with the
$z$ axis in the $yz$ plane.
This leads to vanishing $\mathbb{T}_x$, and $\mathbb{T}_y$
components of the Green's function in $F_1$. As discussed earlier,
the macroscopic phase of the left and middle
superconducting leads 
are $\theta_L=\theta_M=0$ while $\theta_R=\pi/2$. The top row of the
figure corresponds to equal-thickness magnets, with
$d_{F1}=d_{F2}=0.4\xi_S$, whereas the bottom row is for the same
parameter set except now $d_{F1}=0.4\xi_S$ and $d_{F2}=0.7\xi_S$. In
both cases, we consider a rather thin $S$ lead namely,
$d_S=0.2\xi_S$. As can be seen, the total current in both $F_1$ and
$F_2$ depends on the magnetization rotation of $F_2$. In the top
row, the supercurrent in both $F_1$ and $F_2$ behave similarly. The
current is positive when the relative magnetizations are in the
parallel state ($\beta_2=0$), and then the current changes direction
in the $F_1$ layer after $\beta_2=\pi/2$, corresponding to
perpendicular relative magnetization directions, and transition to a
$\pi$ state. Turning now to the individual components of the
supercurrent, we see
from the top panel of Fig.~\ref{fig:sfsfs_beta} that the singlet contribution, $I_{S0}$, 
follows some similar trends as the total current, but with different
magnitudes. The average behavior of $I_{S0}$ over the $F$ regions
are shown to both vanish at the same $\beta_2$, indicating
that the singlet part of the total supercurrent changes sign within the magnets.
The possible spin-polarization effects due to the
magnetization misalignment of the two $F$ layers
is revealed in the odd-frequency triplet contributions, 
$I_{Sy}$, and $I_{Sz}$. The plots clearly  demonstrate that the
spin-1 projection of the triplet current, $I_{Sy}$, peaks in $F_2$
when the relative magnetizations are nearly orthogonal
($\beta_2\approx\pi/2$), corresponding to nearly complete
magnetization alignment along $y$. The $I_{Sy}$ component  meanwhile
vanishes in $F_1$ as expected since the magnetization is aligned
with the $z$ quantization axis. The odd-frequency component,
$I_{Sz}$, is typically finite  in both magnets, possessing its
largest value when their relative magnetizations are aligned along
$z$. There is no average current flow when the relative
magnetizations are approximately orthogonal ($\beta_2=\pi/2$).
Therefore, the device may also be viewed as a charge supercurrent
switch controlled by  magnetization orientation.

The bottom set of panels demonstrate that for differing $F$ layer
thicknesses the magnetization rotation can render one part of the
system to be in a $0$-state and the other to be in the $\pi$ state.
This suggests a $0$-$\pi$ spin switch that can arise by simply
rotating the magnetization orientation in one of the ferromagnetic
layers.  Since $h_x=h_y=0$ in $F_1$, we have the $\mathbb{T}_x$ and
$\mathbb{T}_y$ contributions to the supercurrent vanishing (see
Eq.~(\ref{eq:current})). This is consequently also observed
in the averaged equal-spin triplet current, 
$I_{Sy}(x)$. Note that we do not consider 
magnetizations out of plane, and therefore $I_{Sx}(x)$ necessarily vanishes throughout the system.
To pinpoint the precise behavior of the total supercurrent and 
its spatially averaged  triplet components,
it is
insightful
to study their explicit spatial dependence.

In Fig.~\ref{fig:sfsfs_x}, we therefore illustrate the total charge 
supercurrent and its components as a function of position inside the
three-terminal junction. 
Two representative
angles $\beta_2=$ $\pi/8$ and $7\pi/8$ are chosen, and four
different phases of the right superconductor, $\theta_R$, are
considered:
$\pi/8$, $\pi/4$, $3\pi/8$, and $\pi/2$. The total current is 
piecewise constant in each non-superconducting region, reflecting
local charge conservation. The central $S$ region, however, acts as
an external source of Cooper pairs, and thus the charge current in
that region will acquire a position-dependence 
profile. This can be verified by considering the second set of
panels from the left in Fig.~\ref{fig:sfsfs_x}, which depict the
singlet contribution, $I_{S0}(x)$, as a function of position. The
outer $s$-wave superconducting leads combined with the inhomogeneous
magnetization provided by the two $F$ layers, induces odd-frequency
triplet correlations that naturally are location-dependent as well.
This is observed  in the other remaining panels. The middle $S$
layer is void of any equal-spin, odd-frequency triplet correlations
($I_{Sy}$), but is populated with triplet $I_{Sz}$ superconducting
correlations that clearly depend on $\beta_2$  and $\theta_R$.
Examining the panels of Fig.~\ref{fig:sfsfs_x}, it is seen that the
odd-frequency triplet component with nonzero spin projection,
$I_{Sy}(x)$,
vanishes in $S$. 
This is in contrast to the
$I_{Sz}(x)$ contribution, which is constant inside the middle $S$ terminal and equal to 
its value at the $S/F$ interface.
Thus within the middle $S$ lead, the nondecaying odd-frequency triplet component, $I_{Sz}$,
and the singlet component, $I_{S0}$,
have a direct influence on supercurrent control.
For this reason, within the $F_1$ region,
the net supercurrent flow is
seen to be due to the competition
solely between $I_{S0}$ and $I_{Sz}$ (since $I_{Sy}$ vanishes there), and
which sometimes are oppositely directed.
Magnetization rotation in $F_2$ can thus result in
total supercurrent flow there that is opposite to that of $F_1$. 
Consequently the system resides in a composite $0$- and
$\pi$-state. Since the supercurrent is conserved inside the
non-superconducting regions, this also implies that within the central
$S$ layer itself, the total supercurrent must undergo a reversal in
direction. 

\begin{figure}[t!]
\includegraphics[width=8.5cm,height=4.0cm]{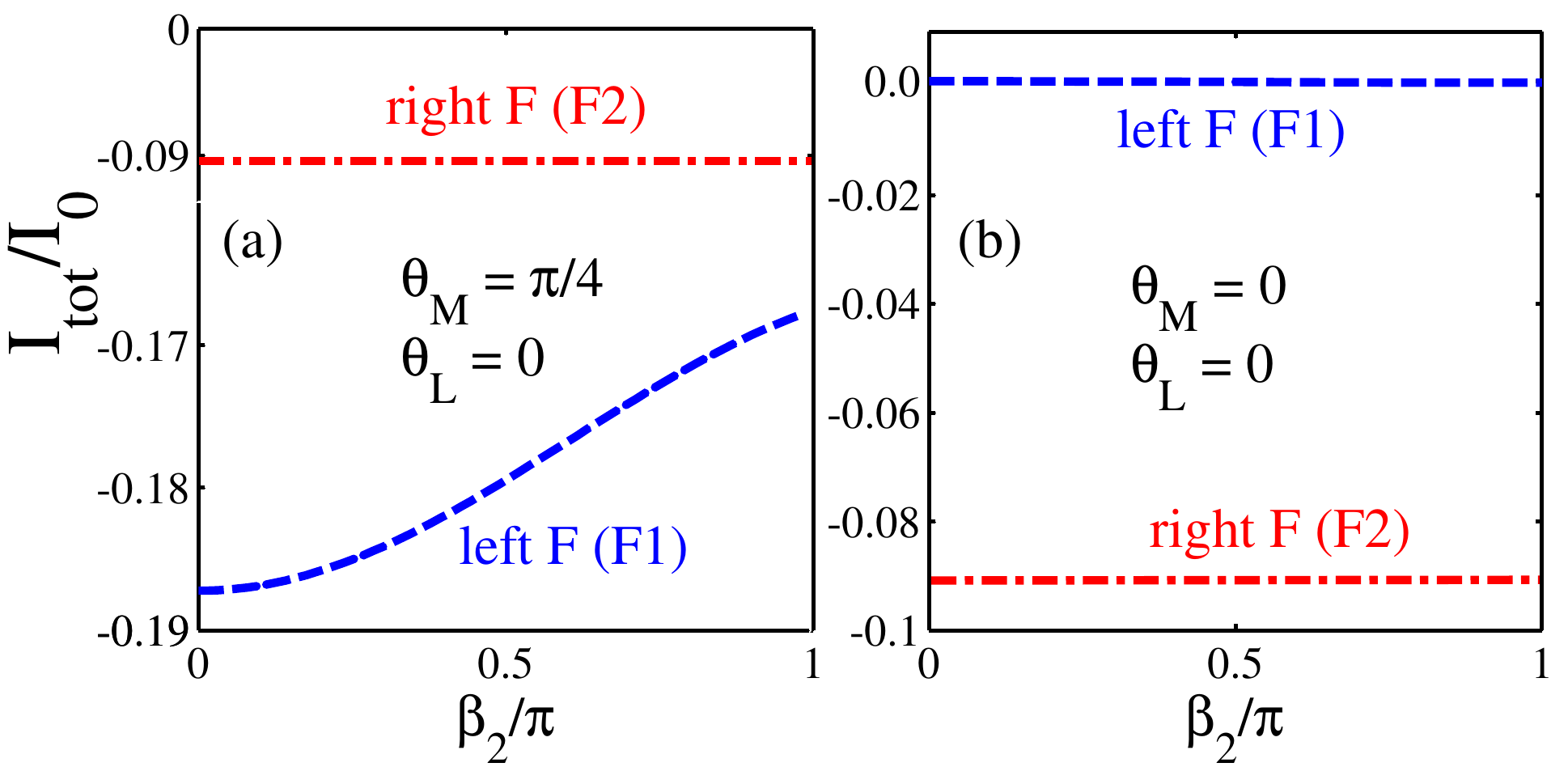}
\caption{\label{fig:pumping} (Color online) Total maximum
supercurrent in each $F$ region against $\beta_2$, the magnetization
orientation of $F_2$  in $SFSFS$ junction. Also, the magnetization
in $F_1$ is fixed along the $z$ axis, i.e., $\beta_1=0$. Here a
thick middle $S$ terminal is considered with
$d_S=4.0\xi_S$ corresponding to several tens of nanometers and 
$d_{F1}=0.4\xi_S$, $d_{F2}=0.7\xi_S$. (a) the macroscopic phase of
the middle $S$ terminal is fixed at $\theta_M=\pi/4$ while in (b)
this quantity is set to zero, $\theta_M=0$. In both cases, 
the macroscopic phase of the left superconducting terminal is
$\theta_L=0$ while $\theta_R$ varies in order to find the maximum
supercurrent.}
\end{figure}

As mentioned above, for large middle $S$ layer widths, $d_S$,
and $\theta_L-\theta_M=0$,  
the outer $S$ terminals should generally become decoupled, making it
impossible to manipulate the current flowing in $F_1$ via
magnetization rotation in $F_2$. By externally tuning the
macroscopic phase of the middle $S$ layer however, the total maximum
charge current in $F_1$ can now be controlled by the
rotation of  magnetization in $F_2$. This is 
illustrated in Fig.~\ref{fig:pumping}, where the total current in
each magnetic region of a $SFSFS$ structure is plotted versus the
magnetization orientation of  $F_2$ layer, $\beta_2$. As
before, the magnetization of $F_1$  is fixed along the
$z$ axis, i.e., $\beta_1=0$. The thicknesses of the
$F$ layers are
$d_{F1}=0.4\xi_S$, and  $d_{F2}=0.7\xi_S$, while a relatively 
thick middle $S$ layer is set to $d_S=4.0\xi_S$.
This
choice of $d_S$
permits an analysis of
the coupling and 
supercurrent roles
of the middle $S$ terminal.
The superconducting phase of the left $S$ terminal is fixed at
$\theta_L=0$, whereas $\theta_R$ varies over $[0,2\pi]$ to
determine the maximum supercurrent flow. In Fig.~\ref{fig:pumping}(a), the macroscopic phase of
the middle $S$ terminal is
equal to $\theta_M=\pi/4$. As seen there, the supercurrent in
$F_2$  is insensitive to magnetization direction, $\beta_2$.
This is consistent with the fact that
charge supercurrent in a
single $SFS$ junction must be constant and independent of magnetization 
rotation. This is contrary to the supercurrent behavior
in $F_1$, where variations
are shown as the magnetization vector rotates.
In a way similar to what was observed in Fig. \ref{fig:sfsfs_x},
the odd-frequency triplet current in $F_2$ is
partially propagated through the middle $S$ electrode into
$F_1$. The transport of these
superconducting correlations into
$F_1$  constitute a coupling
mechanism between the outer $S$ banks.
Therefore,  the odd-frequency triplet correlations
and even-frequency singlet correlations together
result
in changes to the supercurrent in $F_1$  
by  magnetization rotation of $F_2$. 

\begin{figure}[b!]
\includegraphics[width=8.5cm,height=4.0cm]{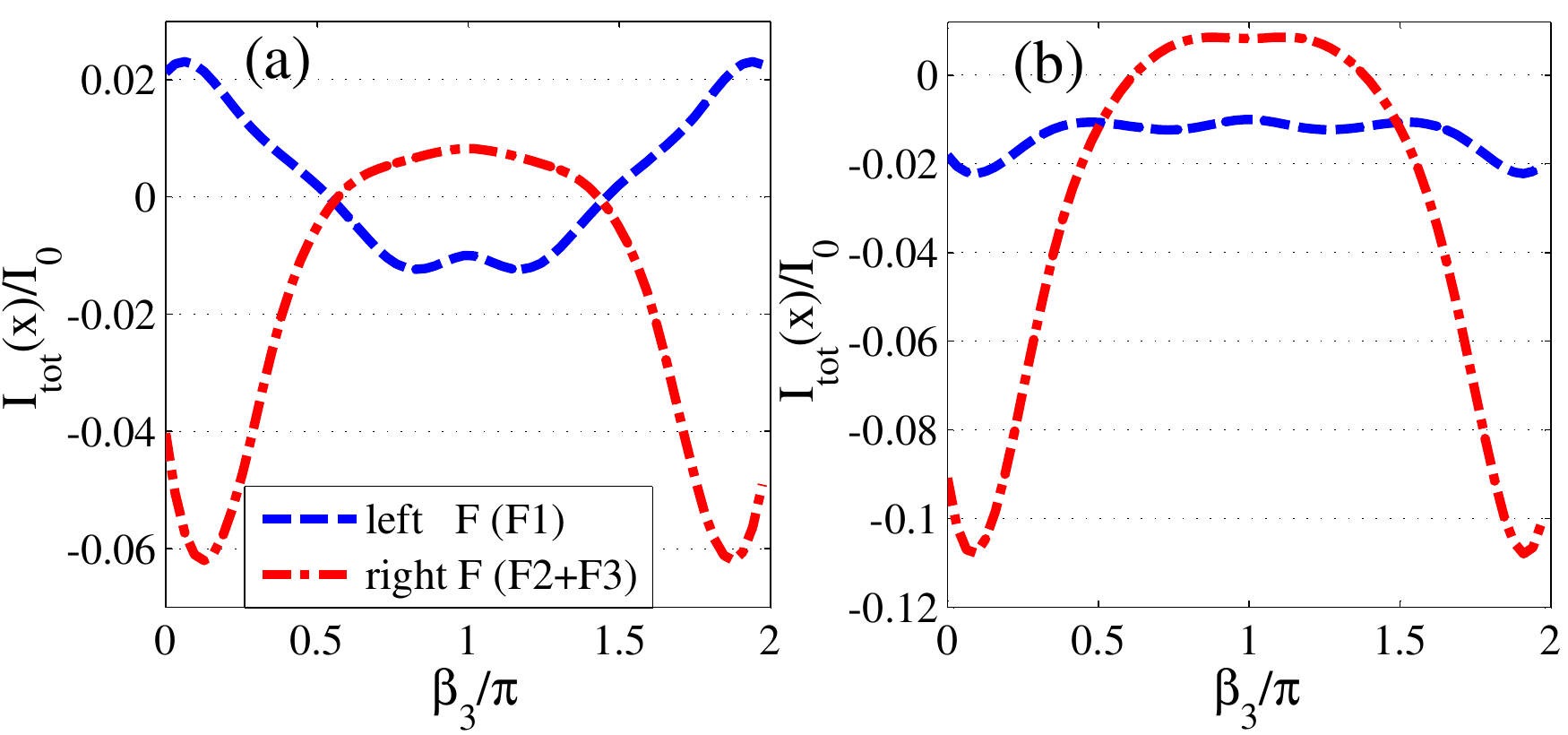}
\caption{\label{fig:sfsffs_beta} (Color online) (a) Supercurrent in
a $SFSFFS$ junction where $\beta_1=0$, $\beta_2=\pi$, and $\beta_3$
is varied.
The geometrical parameters correspond to $d_S=0.2\xi_S$,
$d_{F1}=0.4\xi_S$, and $d_{F2}=d_{F3}=0.3\xi_S$. (b) Josephson
current in the same structure with the same magnetization
orientation, but now with $d_{F2}=0.45\xi_S$,
$d_{F3}=0.15\xi_S$, and unchanged $d_S$.
As done previously, we set $\theta_L=\theta_M=0$ 
and $\theta_R=\pi/2$.
The supercurrent is conserved within each 
magnetic layer (see Figs. \ref{fig:model} and \ref{fig:sfsfs_x}).
Thus the current is the same in the double ferromagnet regions as
clearly seen in Fig. \ref{fig:sfsffs_x}. }
\end{figure}

\begin{figure*}[t]
\includegraphics[width=17.cm,height=7.70cm]{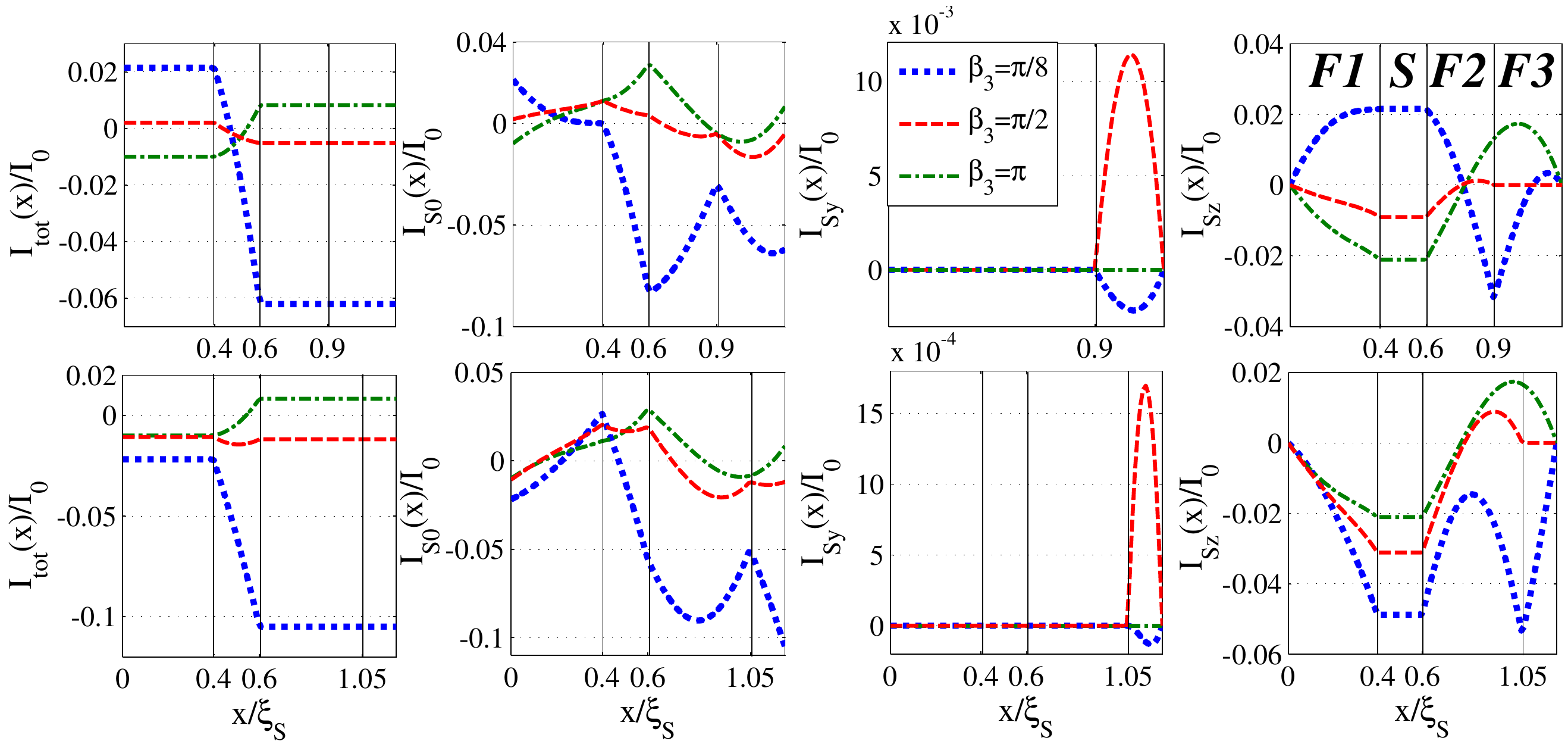}
\caption{\label{fig:sfsffs_x} (Color online) $SFSFFS$ junction (see
Fig.~\ref{fig:model}). The spatial profiles of 
the Josephson supercurrent and its components (singlet ($I_{S0}$) and
odd-frequency triplet ($I_{Sy,z}$)). The two magnetizations in $F_1$
and $F_2$ are fixed in an antiparallel configuration ($\beta_1=0$,
$\beta_2=\pi$) and three different magnetic orientations, $\beta_3$,
are considered for $F_3$.
The macroscopic phases, as before, are set to $\theta_L=\theta_M=0$ 
and $\theta_R=\pi/2$. In all cases, the width of $F_1$ and middle
$S$ electrode are 
fixed at $d_{F1}=0.4\xi_S$ and $d_{S}=0.2\xi_S$, respectively. The
other geometrical parameters correspond to (a) top row:
$d_{F2}=d_{F3}=0.3\xi_S$, and (b) bottom row: $d_{F2}=0.45\xi_S$,
$d_{F3}=0.15\xi_S$. The vertical lines identify the interfaces of
the junction 
among the magnetic and superconducting regions labeled by $F_1$,
$F_2$, $F_3$ and $S$, respectively. }
\end{figure*}
In Fig.~\ref{fig:pumping}(b) we further
explore the proximity effects related to
the width of the middle $S$ terminal, where it
now has a phase of zero ($\theta_M=0$).
It is evident that the
total critical supercurrent in $F_2$  is non-zero and
constant for all values of magnetization orientations,
due to the nonzero phase difference between the middle and 
right $S$ terminals. However, the supercurrent in  $F_1$
vanishes despite a phase difference between
the outer $S$ terminals.
This clearly
demonstrates that 
for sufficiently thick middle $S$ terminals
and proper choice of phase differences,
the outer $S$ electrodes can become decoupled.
We may 
summarize our results using Eq. (\ref{eq:sinusoidal}) in the following way: 
For a supercurrent in  $F_1$ and thick middle $S$
layer, $d_S\gg\xi_S$, we have, $I_1\rightarrow 0$. If we set then
$\varphi_{LM}=0$, no current flows through $F_1$ (see
Fig.~\ref{fig:pumping}(b)). For the case of a thin middle $S$ layer,
and $\varphi_{LM}=0$, the first term in Eq.~(\ref{eq:sinusoidal})
demonstrates that the magnitude of the supercurrent
in $F_1$ is largest when $|\theta_L-\theta_R|=\pi/2$, 
in accordance with Fig.~\ref{fig:sfsfs_beta}. When
$\theta_R-\theta_M \neq 0$, and $\theta_L-\theta_M \neq 0$,
the third term in Eq.~(\ref{eq:sinusoidal})
contributes to the generation of supercurrent
(in addition to the non-zero second
term).
Interestingly, this coupling
term involves the product of $\sin\varphi_{RM}$ and
$\sin\varphi_{LM}$, which for our parameters,
and $S$ layers a few $\xi_S$ thick,
is the dominant term in Eq.~(\ref{eq:sinusoidal}).
In this case, the coupling term 
reveals itself only in 
$F_1$ where there is a negative phase gradient, from right to left
(see Fig.
\ref{fig:pumping}(a)). 
Thus, one may conclude that, e.g., if we set
$\theta_L<\theta_M<\theta_R$, the middle $S$ layer 
mediates supercurrent flow through $F_1$ via
magnetization rotation in $F_2$.

It is important to note that we have directly solved the
Usadel equations (Eqs. (\ref{Linearized Usadel Eq.})) in the $F$
regions and the middle $S$ electrode together with appropriate boundary
conditions Eqs. (\ref{bc_F1}) and (\ref{bc_F2}). This way, we match the
Green's function at the
interfaces and thus the interaction of adjacent $F$ regions for
relatively thin middle $S$ electrodes can be fully accounted for.
If on the other hand, one uses the bulk solution given by Eq. (\ref{Eq:bulk_solution})
for the middle $S$
electrode instead of solving the appropriate equations, the middle $S$ region  
prohibits any transport between the adjacent $F$ regions similar to the
$d_S\gg \xi_S$ regime discussed earlier.

We now introduce additional magnetic inhomogeneity 
into the system by considering a more complicated $SFSFFS$ structure
as shown in part $ii)$ of Fig.~\ref{fig:model}. The maximum value of
the total charge current is shown in Fig.~\ref{fig:sfsffs_beta} over
two regions: The $F_1$ region, and the $F_2$$/$$F_3$ double layer.
Here the geometrical parameters correspond to
$d_{F2}=d_{F3}=0.3\xi_S$ (panel (a)) and $d_{F2}=0.45\xi_S,
d_{F3}=0.15\xi_S$ (panel (b)). In both cases, $d_{F1}=0.4\xi_S$, and
$d_S=0.2\xi_S$ (this value of $d_S$ is correspond to the thin middle $S$ electrode discussed earlier). 
We also set the phases, $\theta_L=\theta_M=0$, and
$\theta_R=\pi/2$. We vary the magnetization of the $F_3$ layer,
$\beta_3$, while fixing $\beta_1=0$, and $\beta_2=\pi$. Thus the magnetization in
$F_1$
is oriented along $z$, antiparallel to $F_2$.
As can be seen, the supercurrent direction and magnitude in each $F$
region can be controlled by the magnetization orientation in $F_3$.
In panel (a) the total current in each ferromagnet is directed
oppositely over the whole angular range of $\beta_3$, except for
$\beta_3\approx\pi/2$ corresponding to magnetizations nearly
orthogonal to the other two. In this case, there is a vanishing of
the supercurrent in all $F$ regions.
The reversal of supercurrent direction in the $F$ segments
upon varying the magnetization orientation in the $F_3$ region, is in
stark contrast to the findings of the previous $SFSFS$ case (bottom row
of Fig. \ref{fig:sfsfs_beta}), where similar geometrical
parameters were used. There the supercurrent changes direction in
$F_1$  whereas it remains unchanged in $F_2$ by varying $\beta_2$ for the
$d_{F1}=0.4\xi_S$, and $d_{F2}=0.7\xi_S$ case (where
$d_{F2}>d_{F1}$).
However, for the $SFSFFS$ junction
with equal $F_2$, $F_3$ thicknesses (with $d_{F2}+d_{F3}>d_{F1}$),
and parameters given in Fig.~\ref{fig:sfsffs_beta}(a),
we find the system
has the $0$ and $\pi$ states coexisting
over nearly the whole
angular range of $\beta_3$.
An exception occurs
near $\beta_3 \approx \pi/2$ and $3\pi/2$,  where
the supercurrent vanishes.
It is also seen in  Fig.~\ref{fig:sfsffs_beta} that the $0$- and
$\pi$-states exchange locations upon varying the magnetization
rotation $\beta_3$. In other words, the coexistence of $0$- and
$\pi$-states in the $SFSFS$ junction is now enhanced in the $SFSFFS$
case. This interesting effect in $SFSFFS$ junctions tends to wash
out when $d_{F2}\neq d_{F3}$ (see Fig.~\ref{fig:sfsffs_beta}(b)).
It is apparent
that
the transport characteristics of $SFSFFS$ Josephson junctions
can be highly sensitive to the
geometrical parameters and magnetization patterns.
Clearly, the addition of the $F_3$ layer
increases the possible tunable parameters, e.g., its width and
magnetization orientation, so that more possibilities
arise for spin switching and supercurrent control.

The behavior of the supercurrent as a function of magnetization
variation in $F_3$ (Fig.~\ref{fig:sfsffs_beta}) is consistent with
the local spatial profile of the total supercurrent and its even and
odd frequency components, exhibited in Fig.~\ref{fig:sfsffs_x}. In
particular, the top panels illustrate that the total supercurrent is
positive in throughout the $F_1$ region and then as $\beta_3$
increases, the current switches direction, becoming negative. The
reverse trends are observed in the remaining ferromagnet regions,
$F_2$ and $F_3$, in accordance with Fig.~\ref{fig:sfsffs_beta}(a).
The valve effect is clearly identified for the perpendicular
magnetic configuration ($\beta_3$), where the supercurrent nearly
vanishes throughout the entire system. Turning now to the individual
components of the total supercurrent, we see that although the
current must be uniform in the $F$ regions, the even and odd
frequency contributions can have complicated spatial behavior. The
spin-1 triplet component, $I_{Sy}$, is shown to vanish when all
three magnetizations are collinear, which occurs when $\beta_3=\pi$.
As expected, it vanishes in the $F_1$ and $F_2$ regions for all
$\beta_3$ since the relative magnetization there is always
collinear. When the ferromagnet layers have magnetizations that are
no longer collinear, spin-1 triplet correlations can be generated,
which is largest  in $F_3$ for $\beta_3= \pi/2$. \cite{Houzet1} On
the other hand, the triplet component, $|I_{Sz}|$ is largest when
the magnetization angle corresponds to values closer to the $z$ spin
quantization axis, which in this case are $\beta_3=\pi/8$ and $\pi$.
Although similar trends are observed when considering unequal $F$
widths (bottom row), the configuration involving larger $d_{F2}$ and
e.g., $\beta_3=\pi/8$ permits $I_{Sz}$ to establish a maximum in
$F_2$ and subsequent decline towards the middle $S$, so that there
is a greater contribution to negative total current flow compared to
the symmetric case in the row above.

\section{Conclusions}\label{conclusion}

In conclusion, we have considered $SFSFS$ and $SFSFFS$ systems which
have been recently realized experimentally and are expected to have
potential applications in the next generation of memories and quantum
computers.\cite{Baibich,eschrigh1,Grunberg,Ioffe,ryaz1,ryaz2,ryaz3,ryaz4,pugach1,giaz1,giaz2,alidoust3}
We have considered the broadly accessible diffusive regime,
which is applicable to many experimental conditions. Using the
Keldysh-Usadel quasiclassical method, we demonstrated that in $SFSFS$
and $SFSFFS$ systems the behavior of the
supercurrent in a given $SFS$ segment can remarkably
be controlled simply
by the magnetization orientation in
the other ferromagnetic regions.
We have shown that
$0$-$\pi$ state profiles of each junction segment is controllable by means of
magnetization rotation. Particularly, the magnetization
rotation can render one part of the junction to be in a $0$ state
while the other can be in a $\pi$ state. In other words, the system can be
in both a $0$ and $\pi$ state configuration: a
three-terminal $0$-$\pi$ 
spin switch. We have investigated the current-phase relations in such
structures  numerically, and formulated our findings.
Our results revealed that a relatively thick middle $S$ electrode
can act as an external
source of supercurrent
or can effectively limit the spin-tuned transport through the system,
depending on the macroscopic superconducting phases.
We have
analyzed the origin of such aspects by decomposing the total
supercurrent into its even-frequency singlet and odd-frequency
triplet
components.
We have shown that the
triplet correlations  propagate
through the middle superconductor
terminal  without any decline in their amplitude. This is
suggestive of a superconducting spin-switch with controllable
charge supercurrent using the magnetization rotation of a
ferromagnetic layer constituting  the $SFSFS$ and $SFSFFS$ systems.

\acknowledgments

M.A. would like to thank G. Sewell for valuable discussions in
numerical parts of this work. K.H. is supported in part by ONR and
by a grant of supercomputer resources provided by the DOD HPCMP.

\section{Pauli Matrices}
In Sec.~\ref{section:theory} we introduced the
the $2\times 2$ Pauli matrices in spin space. 
They are denoted by $\vec{\tau}=(\tau_1,
\tau_2, \tau_3)$, and given by,
\begin{align}
&\tau_1 = \begin{pmatrix}
0 & 1\\
1 & 0\\
\end{pmatrix},\;
\tau_2 = \begin{pmatrix}
0 & -i\\
i & 0\\
\end{pmatrix},\;
\tau_3 = \begin{pmatrix}
1& 0\\
0& -1\\
\end{pmatrix}.\nonumber\\*
&\text{We also introduced the $4\times 4$ matrices
$\vec{\hat{\rho}}=(\hat{\rho}_1, \hat{\rho}_2, \hat{\rho}_3)$:}\nonumber\\\nonumber &\hat{\rho}_1 =
\begin{pmatrix}
0 & \tau_1\\
\tau_1 & 0 \\
\end{pmatrix},\;
\hat{\rho}_2 =  \begin{pmatrix}
0 & -i\tau_1\\
i\tau_1 & 0 \\
\end{pmatrix},\;
\hat{\rho}_3 = \begin{pmatrix}
\tau_0 & 0\\
0 & -\tau_0  \\
\end{pmatrix}.
\end{align}
To simplify expressions,
it is also convenient to use the following definitions: 
\begin{align}
 \tau_0 = \begin{pmatrix}
1 & 0\\
0 & 1\\
\end{pmatrix},\;\hat{\rho}_0 = \begin{pmatrix}
\tau_0 & 0\\
0 & \tau_0 \\
\end{pmatrix}.\nonumber
\end{align}

\end{document}